\documentclass[aps,prl,twocolumn,superscriptaddress,showpacs,floatfix,nofootinbib]{revtex4-1}
\usepackage{amsmath,graphicx}
\newcommand{\avgev}[1]{\left\langle{#1}\right\rangle}
\newcommand{\avgevvn}[1]{\left\langle{#1}\right\rangle_{|v_n}}

\newcommand{\Qf}[2]{\frac{Q_{#1}^{#2}}{|Q_{#1}^{#2}|}}
\newcommand{\Qfs}[2]{\frac{Q_{#1}^{*#2}}{|Q_{#1}^{#2}|}}
\begin{document}

\title{Event-plane correlators}

\author{Rajeev S. Bhalerao}
\affiliation{Department of Theoretical Physics, Tata Institute of Fundamental Research,
Homi Bhabha Road, Mumbai 400005, India}
\author{Jean-Yves Ollitrault}
\affiliation{
CNRS, URA2306, IPhT, Institut de physique theorique de Saclay, F-91191
Gif-sur-Yvette, France} 
\author{Subrata Pal}
\affiliation{Department of Nuclear and Atomic Physics, Tata Institute of Fundamental Research, 
Homi Bhabha Road, Mumbai 400005, India}
\date{\today}

\begin{abstract}
Correlators between event planes of different harmonics in
relativistic heavy-ion collisions have the potential to provide
crucial information on the initial state of the matter formed in these
collisions. We present a new procedure for analyzing such correlators,
which is less demanding in terms of detector acceptance than the one
used recently by the ATLAS collaboration to measure various two-plane
and three-plane correlators in Pb-Pb collisions at LHC.  It can also
be used unambiguously for quantitative comparison between theory and
data. We use this procedure to carry out realistic simulations within
the transport model AMPT. Our theoretical results are in excellent
agreement with the ATLAS data, in contrast with previous hydrodynamic
calculations which only achieved qualitative agreement.  We present
predictions for new correlators, in particular four-plane correlators,
which can easily be analyzed with our new method.
\end{abstract}


\maketitle

$Introduction.-$High-energy heavy-ion collision experiments at the
Relativistic Heavy-Ion Collider (RHIC), BNL and the Large Hadron
Collider (LHC), CERN have firmly established the formation of strongly
interacting matter which exhibits a strong collective
flow~\cite{Heinz:2013th,Gale:2013da,Ollitrault:1992bk}. This not only suggests
that the matter formed is close to local thermal equilibrium but also
provides a window to the initial state of the fireball immediately
after the collision. Collective flow in the plane transverse to the
beam axis is typically measured in terms of two-particle angular
correlations~\cite{Adcox:2002ms,Aamodt:2011by,Chatrchyan:2012wg,ATLAS:2012at}
and its small anisotropies captured in its Fourier
harmonics~\cite{Voloshin:2008dg,Alver:2010gr,Mishra:2007tw,Alver:2006wh}.
Recently, a new tool, namely correlations among event planes
corresponding to different
harmonics~\cite{Bhalerao:2011yg,ALICE:2011ab,Jia:2012sa}, is emerging
with a promise to throw additional light on the initial-state
phenomena.

Pair correlations (i.e., the single-particle
anisotropic flow $v_n$ extracted from two-particle correlations) are
reasonably well understood~\cite{Voloshin:2008dg}. 
Event-plane correlators represent higher-order correlations, involving at
least three particles. Such higher-order correlations thus open
a new direction in heavy-ion physics, much in the same way as studies
of non-Gaussian fluctuations~\cite{Yadav:2007yy} in the early
Universe. They bring in a large number of new observables which
provide new, detailed insight into the hydrodynamic response and on the
initial stage, and will significantly improve our understanding of
heavy-ion collisions.

The ATLAS collaboration~\cite{Jia:2012sa} has recently released
preliminary measurements of eight two-plane correlators (e.g., the
correlation between the second and fourth harmonics) and six
three-plane correlators (e.g., the mixed correlation between second,
third and fifth harmonics) in Pb-Pb collisions at
$\sqrt{s_{NN}}=2.76$~TeV, as a function of the
centrality of the collision.  
These correlators are qualitatively understood within
event-by-event hydrodynamics~\cite{Qiu:2012uy} and provide new
insight~\cite{Teaney:2012gu} into the interplay between the linear and
nonlinear~\cite{Borghini:2005kd} hydrodynamic
response~\cite{Gardim:2011xv,Teaney:2012ke} to the initial density
profile.

The ATLAS analysis is very demanding in terms of detector acceptance:
each event plane is determined in a separate pseudorapidity window;
windows must be pairwise separated by gaps in order to suppress
nonflow correlations; finally, each window must be wide enough to
achieve a significant resolution in every Fourier harmonic up to order
six.

In this article, we show that the analysis can be done more simply, and
that even three- and four-plane correlators can be safely analyzed
with just two symmetric pseudorapidity windows, in the same way as
two-plane correlators. The analysis can thus be performed by other
experiments with smaller detector acceptance. We also explain in
detail how to generalize the scalar-product method used earlier for
two-particle flow analysis ~\cite{Adler:2002pu}, to mixed
correlations, so as to eliminate the ambiguity brought about by the
event-plane method~\cite{Luzum:2012da}.

We carry out realistic simulations within a multiphase transport
(AMPT) model \cite{Lin:2004en}. Results are for the first time in
quantitative agreement with the ATLAS measurements. In addition, we
present predictions for several new correlators, in particular 
{\it four-plane\/} correlators.

$Method.-$The azimuthal distribution of outgoing particles is
decomposed in Fourier harmonics in each collision event.  The flow
vector in harmonic $n$ is defined as~\cite{Poskanzer:1998yz},
\begin{equation}
\label{defqcomplex}
Q_n 
= |Q_n| e^{in\Psi_n}
\equiv\frac{1}{N} \sum_j e^{in\phi_j},
\end{equation} 
where we have used a complex notation and the sum runs over $N$
particles seen in a reference detector, and $\phi_j$ are their
azimuthal angles. $\Psi_n$ is dubbed the event-plane angle in harmonic
$n$. Note that $Q_{-n}=Q_n^*$.

The idea of ATLAS~\cite{Jia:2012sa} is to measure angular correlations
between different harmonics, that is, between $\Psi_n$ and $\Psi_m$
with $n\not=m$. We take as an illustration the correlation between the
2nd and 4th harmonic planes $\Psi_2$ and $\Psi_4$. In order to ensure
that the measured correlation is due to collective
flow~\cite{Luzum:2010sp}, one measures $\Psi_4$ and $\Psi_2$ in
different parts of the detector (``subevents'') $A$ and $B$ separated
by a gap in pseudorapidity (i.e., polar angle). The simplest
observable one can think of is $\avgev{\cos(4(\Psi_{4A}-\Psi_{2B}))}$.
Since, however, $\Psi_{4A}$ and $\Psi_{2B}$ have statistical
fluctuations due to the finite number of particles in each subevent,
one corrects for these statistical fluctuations by a so-called
``resolution'' correction~\cite{Poskanzer:1998yz}. The quantity
measured by ATLAS can be written explicitly as
\begin{equation}
\label{atlas42}
\avgev{\cos 4 (\Phi_2 - \Phi_4)}
\equiv \frac
{\avgev{\Qf{2A}{2}
\Qfs{4B}{}}}
{\sqrt{\avgev{\Qf{4A}{}\Qfs{4B}{}}}\sqrt{\avgev{
\Qf{2A}{2}\Qfs{2B}{2}
}}},
\end{equation}
where we have used the same notation as ATLAS on the left-hand side,
and angular brackets on the right-hand side denote the real part of
the average over events.  The numerator is
$\avgev{\cos(4(\Psi_{4A}-\Psi_{2B}))}$ and the denominator is the resolution
correction \cite{footnote}.

As shown in~\cite{Luzum:2012da}, Eq.~(\ref{atlas42}) represents an
ambiguous measurement, in the sense that the denominator does not
properly correct for the resolution when flow fluctuations are
present~\cite{Alver:2008zza,Ollitrault:2009ie}. If one repeats the
analysis with a lower resolution 
(typically, smaller subevents $A$ and $B$), both the numerator and the
denominator decrease, but the denominator decreases faster than the
numerator, resulting in an increase of the signal.  The difference
between the low-resolution limit and the high-resolution limit can be
as large as 50\% for the above correlation~\cite{Luzum:2012da}. In
practice, experimental values are likely to be close to the
low-resolution limit, due to the poor resolution in higher
harmonics. However, the value calculated in
hydrodynamics~\cite{Qiu:2012uy,Teaney:2012gu} is the high-resolution
limit, so that a quantitative comparison between theory and data is
impossible at present.

These ambiguities can easily be removed. One simply replaces 
Eq.~(\ref{atlas42})  with~\cite{Luzum:2012da}: 
\begin{equation}
\label{mixedsp}
c\{2,2,-4\}\equiv
\frac
{\avgev{Q_{2A}^2Q^*_{4B}}}
{\sqrt{\avgev{Q_{4A}Q^*_{4B}}}\sqrt{\avgev{
Q_{2A}^2Q_{2B}^{*2}
}}},
\end{equation}
where the notation on the left-hand side~\cite{Bhalerao:2011yg}
expresses that the numerator involves two particles in harmonic $2$
(i.e., two factors of $Q_2$) and one particle in harmonic $-4$ (one
factor of $Q^*_4$).  Equations (\ref{atlas42}) and (\ref{mixedsp}) are
referred to as ``event-plane'' (EP) method and ``scalar product'' (SP)
method ~\cite{Adler:2002pu}, respectively.

The only difference is that the scalar-product method retains the
information on the length of $Q_n$. We now explain why this simple 
modification ensures that the right-hand side of Eq.~(\ref{mixedsp})
is a well-behaved flow observable, unlike (\ref{atlas42}). 
The flow picture is a model in which particles are emitted randomly and 
independently according to some underlying probability distribution in each 
event~\cite{Luzum:2011mm}. The azimuthal probability distribution is written as
\begin{equation}
\label{defvn}
P(\phi)=\frac{1}{2\pi}\sum_{n=-\infty}^{+\infty} v_n e^{in\Phi_n} e^{-in\phi},
\end{equation}
where $v_n$ is the anisotropic flow in harmonic $n$ and $\Phi_n$ the corresponding 
reference direction~\cite{Voloshin:1994mz,Luzum:2011mm} (we use the convention $v_0=1$).
The average over events can be performed in two steps, by first averaging
over events with the same $P(\phi)$, 
then averaging over values of $v_n$ and $\Phi_n$~\cite{Luzum:2012da}:
\begin{equation}
\label{2steps}
\avgev{\ldots} = \avgev{\avgevvn{\ldots}}.
\end{equation}
Equations~(\ref{defqcomplex}) and (\ref{defvn}) yield 
$\avgevvn{Q_n}=v_n e^{in\Phi_n}$.
Now, the crucial point is that Eq.~(\ref{mixedsp}) only involves  products of $Q_n$ vectors. 
Since particles are emitted independently, 
the expectation value of a product is the product of expectation values, therefore
\begin{eqnarray}
\label{eachterm}
\avgevvn{Q_{2A}^2Q^*_{4B}}&=& (v_2)^2 v_4 e^{4i(\Phi_2-\Phi_4)},   \cr
\avgevvn{Q_{4A}Q^*_{4B}}&=& (v_4)^2,   \cr
\avgevvn{Q_{2A}^2Q_{2B}^{*2}}&=&  (v_2)^4.
\end{eqnarray}
These equations show that each of the quantities entering Eq.~(\ref{mixedsp}) is a 
well-defined flow observable. 

We now explain why we choose to measure the particular combination in the 
right-hand side of Eq.~(\ref{mixedsp}).
The three observables in Eq.~(\ref{eachterm}) are sensitive to the magnitudes of $v_2$ and $v_4$. 
By taking the ratio in Eq.~(\ref{mixedsp}), one singles out the angular correlation,
in the sense that if one multiplies $v_2$ and/or $v_4$ by a constant, the right-hand 
side of Eq.~(\ref{mixedsp}) is unchanged. 
We therefore call this quantity an ``event-plane correlator'', even though it involves the 
statistics of $v_2$ and $v_4$, and not just the underlying directions $\Phi_2$ and $\Phi_4$. 
Equations~(\ref{2steps}) and (\ref{eachterm}) predict that the right-hand side of Eq.~(\ref{mixedsp}) should 
lie between -1 and 1 by the Cauchy-Schwarz inequality~\cite{Gardim:2012im}.

It can be shown that
Eq.~(\ref{mixedsp}) coincides with Eq.~(\ref{atlas42}) in the limit of
low resolution~\cite{Luzum:2012da} 
but, as shown above, it is independent of the experimental resolution,
unlike Eq.~(\ref{atlas42}), and thus yields an
unambiguous measurement.  The only price to pay for the loss of
ambiguity is a larger statistical error. 

Note that the two-plane correlator defined by Eq.~(\ref{mixedsp}) 
actually involves a {\it three\/}-particle
correlation~\cite{Borghini:2001vi}. This is easily seen by inserting
the definition of $Q_n$, Eq.~(\ref{defqcomplex}), into the numerator
of Eq.~(\ref{mixedsp}). One obtains
\begin{equation}
\label{decomposition}
Q_{2A}^2Q^*_{4B}=\frac{1}{N_A^2N_B}\sum_{j\in A}\sum_{k\in
  A}\sum_{l\in B}
e^{2i\phi_j+2i\phi_k-4i\phi_l},  
\end{equation}
that is, an average of $e^{2i\phi_j+2i\phi_k-4i\phi_l}$ over all
triplets of particles. 

In order to guarantee that correlations are dominated by flow, one
should in principle implement pseudorapidity gaps between {\it all\/}
particles. In Eq.~(\ref{decomposition}), however, there is no
pseudorapidity gap between $j$ and $k$, and there are even
self-correlation terms $j=k$.  Therefore, the procedure followed by
ATLAS does not, strictly speaking, avoid self correlations.

We now explain why the contribution of nonflow correlations and
self-correlations to Eq.~(\ref{decomposition}) is small, thus
justifying the procedure. 
It is generally accepted that long-range correlations are dominated 
by flow~\cite{Luzum:2010sp} or, equivalently, that nonflow correlations 
are short range (with the exception of global momentum conservation, 
which only contributes to the first Fourier harmonic~\cite{Borghini:2000cm}).
Short-range correlations are effects which correlate a particle with its neighbors
with some probability, so that such correlations are typically a small fraction of 
self-correlations~\cite{Bhalerao:2003xf}. 
We thus evaluate the contribution of self-correlations to Eq.~(\ref{decomposition}). 
The thumb rule is that each factor of $e^{in\phi}$ gives a contribution of order $v_n$.
Thus the right-hand side of
Eq.~(\ref{decomposition}) is typically of order $v_4(v_2)^2$.
Self-correlations terms $j=k$ give a contribution of order
$(v_4)^2/N_A$. Since $v_4\sim (v_2)^2$~\cite{Adare:2010ux}, the
relative magnitude of self-correlations is of order $1/N_A$, which is
small. Note that self-correlations can also be removed explicitly if particles 
are seen individually~\cite{Bilandzic:2010jr}. 

The discussion can be repeated for the three-plane correlation between
harmonics 2, 3 and 5. ATLAS measures the three harmonics in three
separate subevents $A$, $B$ and $C$. We now show how the analysis can
be done with just two subevents $A$ and $B$. In close analogy with
(\ref{mixedsp}), we define
\begin{equation}
\label{mixedsp235}
c\{2,3,-5\}\equiv
\frac
{\avgev{Q_{2A}Q_{3A}Q^*_{5B}}}
{\sqrt{\avgev{Q_{2A}Q^*_{2B}}\avgev{Q_{3A}Q^*_{3B}}\avgev{Q_{5A}Q^*_{5B}}
}},
\end{equation}
where the numerator is a mixed correlation between harmonics 2, 3 and
5, and the denominator the corresponding resolution correction in each
harmonic. We now explain why nonflow effects are small, even though
the 2nd and 3rd harmonic are measured in the same subevent $A$. We
decompose the numerator of Eq.~(\ref{mixedsp235}) as
\begin{equation}
\label{decomposition235}
Q_{2A}Q_{3A}Q^*_{5B}=\frac{1}{N_A^2N_B}\sum_{j\in A}\sum_{k\in
  A}\sum_{l\in B}
e^{2i\phi_j+3i\phi_k-5i\phi_l}. 
\end{equation}
Flow typically gives a contribution of order $v_2v_3v_5$. Self
correlations are terms $j=k$, which give a contribution of order
$(v_5)^2/N_A$. Since $v_5\sim v_2v_3$~\cite{Teaney:2012ke}, the
relative order of self-correlations (and therefore of nonflow
correlations) is $1/N_A$, exactly as for the two-plane correlator
$c\{2,2,-4\}$ considered above. Therefore this three-plane correlator
does not require three separate subevents.

This discussion can be easily generalized to all other correlations.
The general prescription for the pseudorapidity gap is that all the
positive harmonics (i.e., all factors of $Q_n$) should go on one side
(e.g., $A$), and all negative harmonics (all factors of $Q^*_n$) on
the other side ($B$).  In this case, self-correlation terms are of
relative order $v_{n+p}/(Nv_nv_p)\sim 1/N$, and nonflow effects can be
safely neglected. Therefore {\it all\/} three-plane correlators
measured by ATLAS can be analyzed with a single pseudorapidity gap.
The analysis can be extended to four-plane correlators and higher.

If $k_n$ denotes the number of particles in harmonic $n$, the
correlator can be generally written as
\begin{equation}
\label{general}
c\{\cdots,\underbrace{n,n,n}_{k_n},\cdots\}\equiv\frac{\avgev{\displaystyle\prod_{n>0}(Q_{nA})^{k_n}\prod_{n<0}(Q_{nB})^{k_n}}}  
{\sqrt{\displaystyle\prod_n \avgev{(Q_{nA})^{k_n}(Q_{nB}^*)^{k_n}}}}. 
\end{equation}
Azimuthal symmetry requires $\sum_n nk_n=0$~\cite{Bhalerao:2011yg}. 
As an example, we write explicitly the formula for the
lowest-order four-plane correlator~\cite{Jia:2012ju}:
\begin{align}
\label{mixedsp2345}
& c\{2,-3,-4,5\}\equiv  \nonumber\\
&\frac
{\avgev{Q_{2A}Q_{5A}Q^*_{3B}Q^*_{4B}}}
{\sqrt{\avgev{Q_{2A}Q^*_{2B}}\avgev{Q_{3A}Q^*_{3B}}\avgev{Q_{4A}Q^*_{4B}}\avgev{Q_{5A}Q^*_{5B}}
}},
\end{align}
for which we give the first quantitative prediction below. 
While two-plane correlations thus defined all lie between $-1$ and 
$+1$ by the Cauchy-Schwarz inequality, there is no such
restriction for three- and four-plane correlators. 
We shall give below an example of a four-plane correlator exceeding
unity in AMPT calculations.

\begin{table}
 \caption{\label{table} 
List of two-, three- and four-plane correlators. For each correlator,
the number in column $n$ indicates the number $k_n$ of particles
involved in harmonic $n$. Asterisks indicate that the particles
actually go into harmonic $-n$ instead of $n$.
The last column is the order of the correlation, i.e., the number of
particles $\sum_n k_n$. 
The first fourteen lines correspond to the correlators measured by ATLAS~\cite{Jia:2012sa}
(correlators in italics are not studied in this paper), 
and the next six lines to new correlators predicted in this paper. 
}
\begin{tabular}{|l|l|l|l|l|l|c|}
\hline
&2&3&4&5&6&Order\\
\hline
$\cos(4(\Phi_2-\Phi_4))$ &2&&1$^*$&&&3\\
$\cos(8(\Phi_2-\Phi_4))$ &4&&2$^*$&&&6\\
$\cos(12(\Phi_2-\Phi_4))$&6&&3$^*$&&&9\\
$\cos(6(\Phi_2-\Phi_3))$ &3&2$^*$&&&&5\\
$\cos(6(\Phi_2-\Phi_6))$ &3&&&&1$^*$&4\\
$\cos(6(\Phi_3-\Phi_6))$ &&2&&&1$^*$&3\\
${\it cos(12(\Phi_3-\Phi_4))}$ &&{\it 4}&{\it 3$^*$}&&&{\it 7}\\
${\it cos(10(\Phi_2-\Phi_5))}$ &{\it 5}&&&{\it 2$^*$}&&{\it 7}\\
$\cos(2\Phi_2+3\Phi_3-5\Phi_5)$ &1&1&&1$^*$&&3\\
$\cos(2\Phi_2+4\Phi_4-6\Phi_6)$ &1&&1&&1$^*$&3\\
$\cos(2\Phi_2-6\Phi_3+4\Phi_4)$ &1&2$^*$&1&&&4\\
$\cos(8\Phi_2-3\Phi_3-5\Phi_5)$ &4&1$^*$&&1$^*$&&6\\
$\it{cos(10\Phi_2-4\Phi_4-6\Phi_6)}$ &{\it 5}&&{\it 1$^*$}&&\it
  {1$^*$}&{\it 7}\\
$\it{cos(10\Phi_2-6\Phi_3-4\Phi_4)}$ &{\it 5}&{\it 2$^*$}&{\it
    1$^*$}&&&{\it 8}\\
\hline
$\cos(2\Phi_2+6\Phi_3-8\Phi_4)$ &1&2&2$^*$&&&5\\
$\cos(3\Phi_3-8\Phi_4+5\Phi_5)$ &&1&2$^*$&1&&4\\
$\cos(9\Phi_3-4\Phi_4-5\Phi_5)$ &&3&1$^*$&1$^*$&&5\\
$\cos(2\Phi_2-3\Phi_3-4\Phi_4+5\Phi_5)$ &1&1$^*$&1$^*$&1&&4\\
$\cos(4\Phi_2-3\Phi_3+4\Phi_4-5\Phi_5)$ &2&1$^*$&1&1$^*$&&5\\
$\cos(6\Phi_2+3\Phi_3-4\Phi_4-5\Phi_5)$ &3&1&1$^*$&1$^*$&&6\\
\hline
\end{tabular}
\end{table}
Table~\ref{table} lists the fourteen correlations studied by ATLAS, as
well as six additional correlations predicted for the first time in
this paper. For clarity, each correlation is written using the same
notation as ATLAS (left column). The next columns list the number of
particles in harmonics 2 to 6. In our notation, the first two
correlators are written as $c\{2,2,-4\}$ and $c\{2,2,2,2,-4,-4\}$,
respectively.  Table~\ref{table} shows that the two-plane correlators
measured by ATLAS actually involve three to nine particles ---never
just two--- yet they can safely be analyzed with a single
pseudorapidity gap, and so can three- and four-plane
correlators. Table~\ref{table} also shows that more particles are
allowed in lower harmonics where the resolution is better.

$Results.-$Calculations are performed using the AMPT model
\cite{Lin:2004en} that consists of four main components: initial
conditions based on Glauber model, parton cascade, hadronization, and
hadron scattering. We employ the version of AMPT with string melting
and parton coalescence for hadronization that describes better the
collective behavior in heavy-ion collisions at RHIC and LHC
\cite{Lin:2004en,Pal:2012gf,Xu:2011jm}.  Flow is mostly produced in
this model by elastic scatterings in the partonic phase. In addition,
the model contains resonance decays and thus includes non-trivial
nonflow effects. The implementation used here is the same as in
Ref.~\cite{Pal:2012gf}, and uses initial conditions from the HIJING
2.0 model~\cite{Deng:2010mv}.  We have checked that it reproduces LHC
data for anisotropic flow ($v_2$ to $v_6$) at all
centralities~\cite{Xu:2011jm,Pal:2012gf,Han:2011iy}.

\begin{figure}[ht]
 \includegraphics[width=\linewidth]{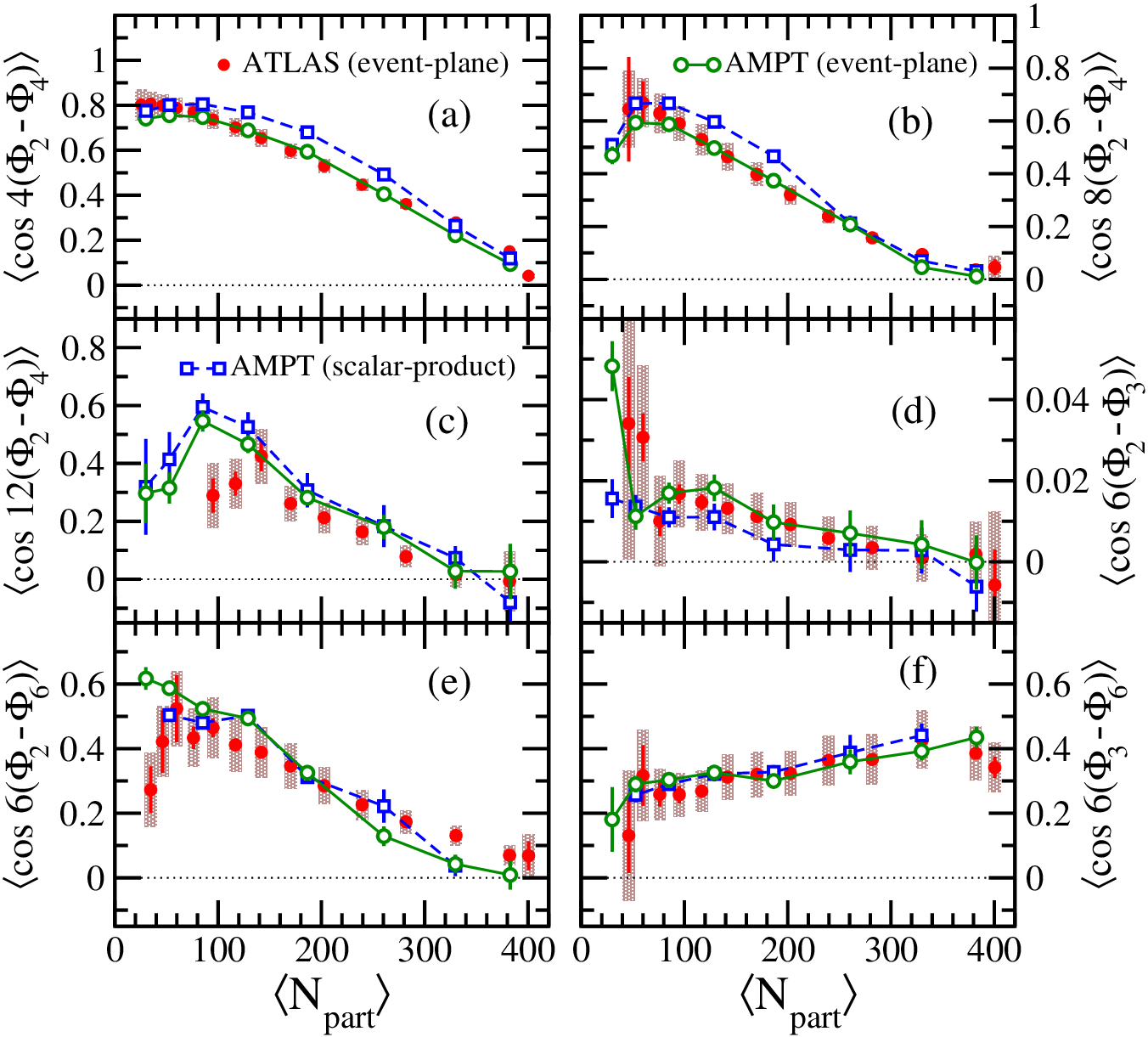}
 \caption{(Color online) Two-plane correlators in the event-plane (open circles) and 
scalar-product (open squares) methods as a function of the number of participants in 
Pb-Pb collisions at $\sqrt{s_{NN}} = 2.76$ TeV in the AMPT model as compared to 
the ATLAS data~\cite{Jia:2012sa} for the event-plane method (solid circles).
We keep the notation of ATLAS $\langle\cos(\cdots)\rangle$ for simplicity, even 
though the measured quantities are not pure event-plane correlators.} 
\label{fig:2p}
\end{figure}

AMPT simulation events can be analyzed with the same methods as actual
experimental events. However, in AMPT, the energy-momentum information
of all final-state particles in an event is available, whereas ATLAS
only detects charged particles with $p_T>0.5$~GeV/c. We, therefore,
use all particles in our analysis which results in a better
resolution.

As explained above, our analysis of event-plane
correlations uses two subevents $A$ and $B$, for which we use 
two symmetric pseudorapidity intervals 
$-4.8<\eta<-0.4$ and $0.4<\eta<4.8$, 
very similar to those by ATLAS for two-plane correlators. 
Each correlator is analyzed using both scalar-product (SP) method
(Eq.~(\ref{general})) and the event-plane (EP) method (in which one
replaces $Q_n$ by $Q_n/|Q_n|$ everywhere in Eq.~(\ref{general})).
Recall that the EP result typically increases (in absolute magnitude)
as the resolution decreases, while the SP result is the limit of low
resolution.  We thus expect that SP is larger than EP.  If the model
is valid, the experimental result (which is obtained with the EP
method, but a lower resolution) should lie somewhere between the two
AMPT predictions.

Figure \ref{fig:2p} displays the results for two-plane correlators.
Figures \ref{fig:2p}(a) and \ref{fig:2p}(b), which have the smallest
error bars, show that the SP method gives larger results than the EP
method, as expected~\cite{Luzum:2012da}. The difference is in the
range (10-15)$\%$. EP results are in perfect agreement with data.
This is in contrast with EP calculations in hydrodynamics, which are
clearly below data~\cite{Qiu:2012uy,Teaney:2012gu}.

\begin{figure}[ht]
 \includegraphics[width=\linewidth]{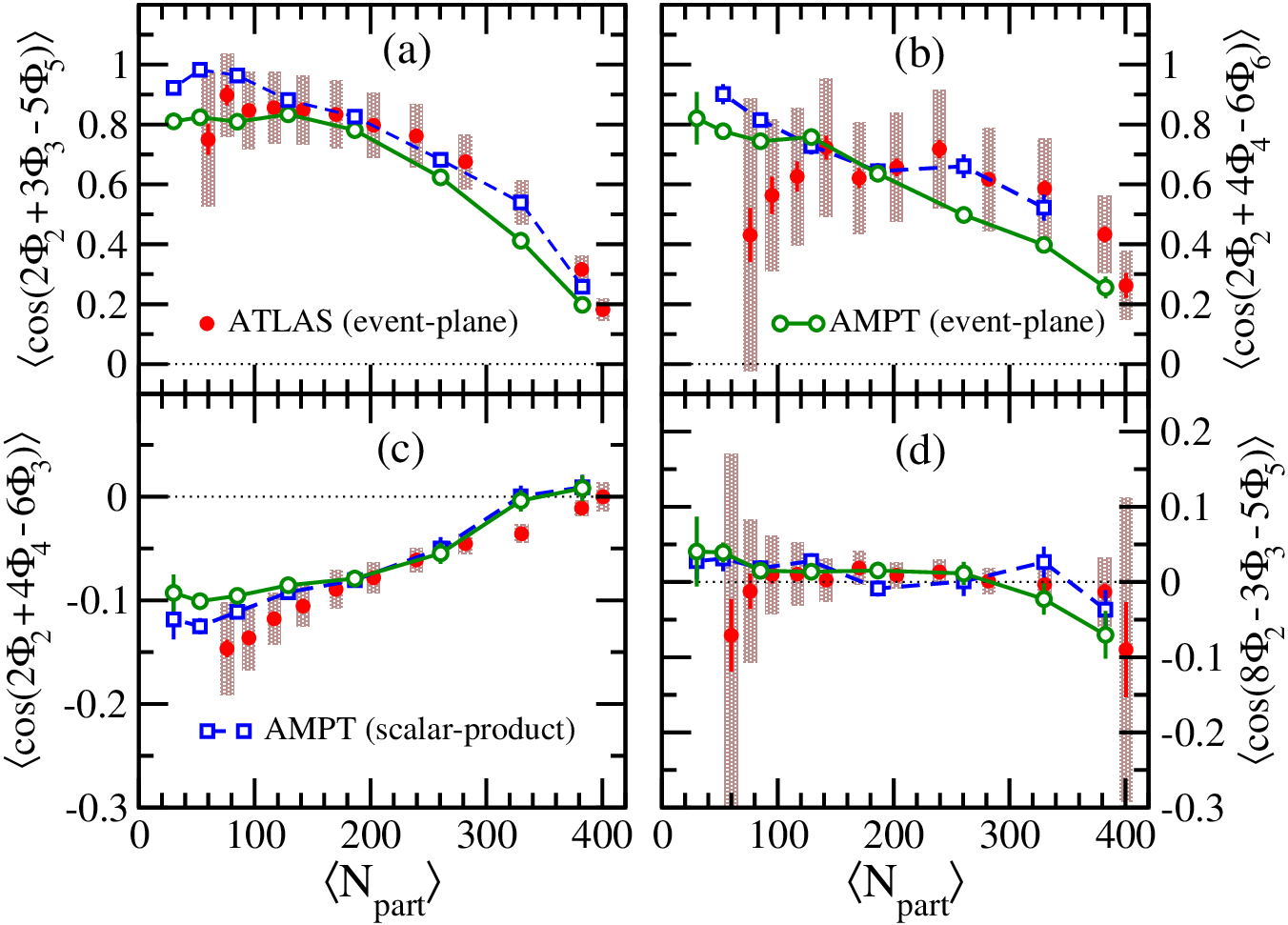}
 \caption{
(Color online) Three-plane correlators. 
Legend is the same as in Fig.~\ref{fig:2p}.}
\label{fig:3p}
\end{figure}

\begin{figure}[b]
 \includegraphics[width=\linewidth]{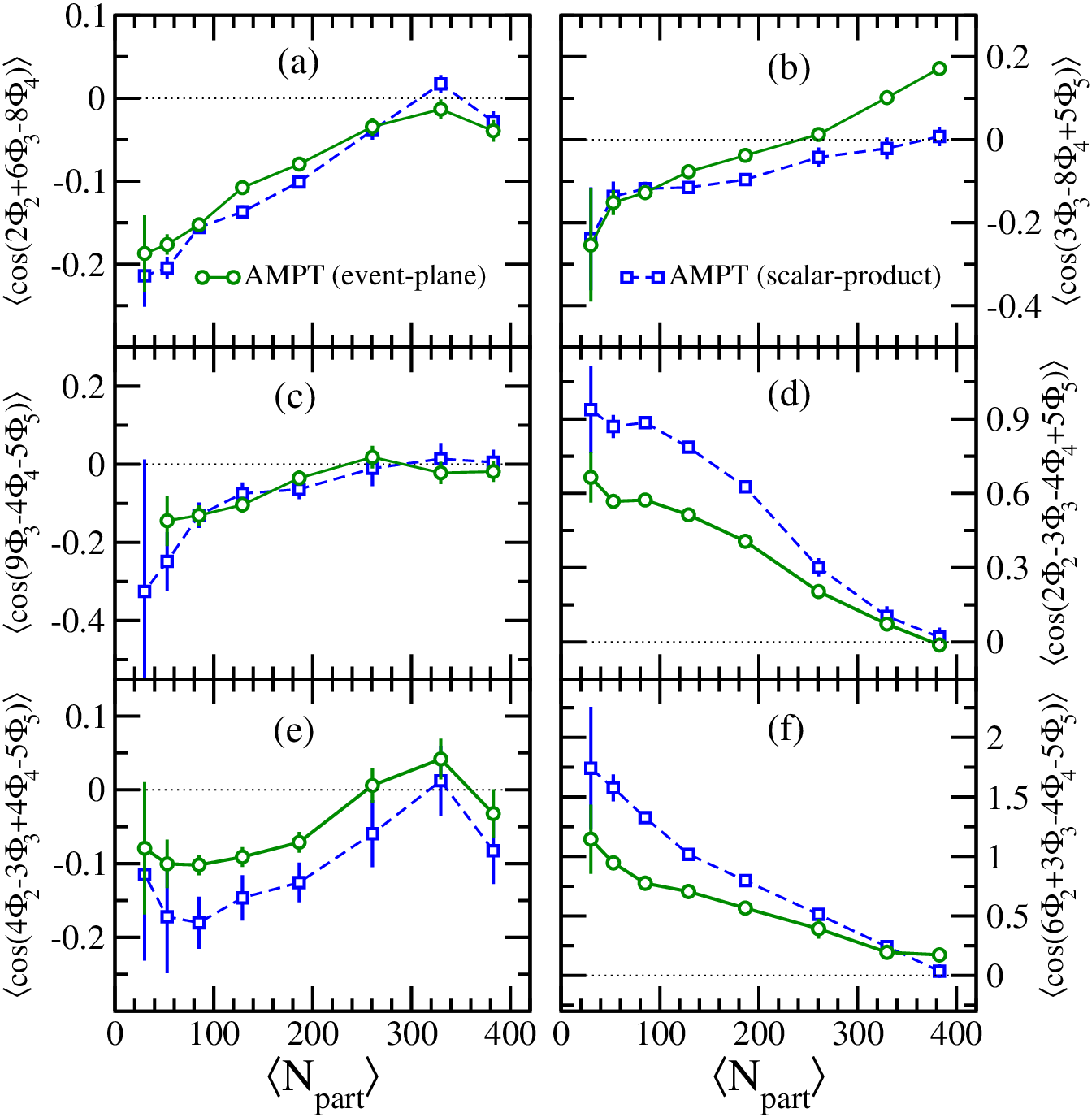}
 \caption{
(Color online) Predictions for new three- and four-plane correlators. 
Legend is the same as in Fig.~\ref{fig:2p}.}
\label{fig:4p}
\end{figure}

The correlations between the second and fourth harmonics
(Figs.~\ref{fig:2p}(a)$-$\ref{fig:2p}(c)) and between the second and
sixth harmonics (Fig.~\ref{fig:2p}(e)) are
understood~\cite{Teaney:2012ke} as coming mostly from the nonlinear
hydrodynamic response, which couples $v_4$ to $(v_2)^2$ and $v_6$ to
$(v_2)^3$. Their increase from central to peripheral collisions is
driven by the increase of $v_2$ itself.  Similarly, the correlations
between harmonics 3 and 6 (Fig.~\ref{fig:2p}(f)) is driven by the
coupling between $v_6$ and $(v_3)^2$: it decreases slightly as the
collision becomes more peripheral, in the same way as $v_3$. Finally,
the correlation between harmonics 2 and 3 (Fig.~\ref{fig:2p}(d)) is an
order of magnitude smaller~\cite{ALICE:2011ab}.  Our calculation
agrees with data in all cases.

Figure \ref{fig:3p} presents AMPT calculations for four three-plane
correlators. Again, calculations are in perfect agreement with
data. $c\{2,3,-5\}$ and $c\{2,4,-6\}$ (Figs.~\ref{fig:3p}(a) and
\ref{fig:3p}(b)) are strong and driven by the nonlinear hydrodynamic
response, in the same way as the correlation between harmonics 2 and
4. The $c\{2,4,-3,-3\}$ correlation has a more subtle
origin~\cite{Teaney:2012gu}, in the sense that it is not simply
generated by the nonlinear response term $v_4\propto (v_2)^2$, but it
is also qualitatively reproduced by event-by-event hydrodynamic
calculations~\cite{Qiu:2012uy}.  Finally, $c\{2,2,2,2,-3,-5\}$ is
compatible with zero.

Figure \ref{fig:4p} presents predictions for three new~\cite{Jia:2012ju}
three-plane correlators (Figs.~\ref{fig:4p}(a)$-$\ref{fig:4p}(c)), as
well as three four-plane correlators
(Figs.~\ref{fig:4p}(d)$-$\ref{fig:4p}(f)).  The magnitude and
centrality dependence of $c\{2,3,3,-4,-4\}$ (Fig.~\ref{fig:4p}(a)) are
similar to those of $c\{2,4,-3,-3\}$ (Fig.~\ref{fig:3p}(c)) and
likewise, this correlator is not simply generated by the $v_4\propto
(v_2)^2$ nonlinear response. Four-plane correlators are much more
sensitive to analysis details than two- or three-plane correlators:
the difference between the scalar-product method and the event-plane
method is roughly 50$\%$. Two of these correlators
(Figs.~\ref{fig:4p}(d) and \ref{fig:4p}(f)) are large, which can again
be understood as an effect of the nonlinear hydrodynamic response
coupling $v_4$ to $(v_2)^2$ and $v_5$ to
$v_2v_3$~\cite{Teaney:2012ke}.  Note that the last four-plane
correlator (Fig.~\ref{fig:4p}(f)) is predicted to exceed unity when
analyzed using the scalar-product method.

$Conclusions.-$We have argued that event-plane correlators in
heavy-ion collisions can be analyzed with just two symmetric
pseudorapidity windows.  We have illustrated the validity of our
approach by analyzing events simulated within the AMPT model which
reproduces for the first time the magnitude and centrality dependence
of the measured correlators in Pb-Pb collisions at LHC.
Much better agreement with data is achieved than in previous
hydrodynamic calculations using the event-plane
method~\cite{Qiu:2012uy,Teaney:2012gu}. This apparent discrepancy
between data and hydrodynamic calculations may simply be due to the
ambiguity of the event-plane method. It will then be resolved once
both experiment and theory switch from the event-plane method to the
scalar-product method. We have presented predictions for new
correlators, in particular large four-plane correlators, which can be
measured in forthcoming analyses. It would be interesting to study the
correlators using the procedure presented here in the event-by-event
hydrodynamical simulations to ascertain the sensitivity to
initial-state models, namely the Monte-Carlo Glauber and the
color-glass-condensate \cite{Gelis:2010nm}.

This work is funded by CEFIPRA under project 4404-2.
JYO acknowledges support by the European Research Council under the
Advanced Investigator Grant ERC-AD-267258.

\end{document}